\documentclass[preprint,amsmath,amssymb,showkeywords,groupedaddress]{revtex4}
\usepackage{graphicx}
\usepackage{dcolumn}
\usepackage{bm}
\usepackage{amsmath}
\usepackage[letterpaper,left=25mm,right=25mm,top=3cm,bottom=2cm]{geometry}
\usepackage{textcomp}
\usepackage{epsfig}
\usepackage{float}
\begin{document}

\title{Entanglement Entropy in an Antiferromagnetic Heisenberg Spin Chain with
Boundary Impurities}

\author{Jie Ren$^1$\footnote{E-mail: jren@cslg.edu.cn}}
\author{Shiqun Zhu$^1$\footnote{Corresponding author, E-mail: szhu@suda.edu.cn
}}
\author{Xiang Hao$^2$}

\affiliation{$^1$School of Physical Science and Technology, Suzhou
University, Suzhou, Jiangsu 215006, People's Republic of China}

\affiliation{$^2$Department of Physics, Suzhou University of Science
and Technology, Suzhou, Jiangsu 215011, People's Republic of China}

\begin{abstract}

The effects of boundary impurities on the entanglement entropy in an
antiferromagnetic Heisenberg opened spin-$1/2$ chain are
investigated. The method of density-matrix renormalization-group is
used to obtain the bipartite entanglement. The entropy increases
when the length of the subsystem increases. It will approach to a
constant when system length is very large. With the same impurity
interaction, qutrit impurities of spin-$1$ can increase the
entanglement entropy. \vspace{1 cm}

\textbf{PACS numbers:} 03.67.Mn, 03.65.Ud, 05.50.+q, 75.10.Jm

\textbf{Key words:} Entropy, Boundary Impurities, DMRG

\end{abstract}

\maketitle

\section{Introduction}

Recently, the entanglement has been recognized as an important
resource of some quantum mechanical phenomena, such as quantum
teleportation, quantum cryptography, quantum computation, and
violation of Bell's inequality \cite{Nielson,Bennett,Murao}. Many
investigations show that entanglement exists naturally in the spin
chain when the temperature of system is at zero. A useful many-body
entanglement measure of a pure state is the von Neumann entanglement
entropy \cite{Bennett1}, which can also quantify quantum phase
transition \cite{Sachdev,Gu1,Anfossi,Lambert,Larsson,Molina,Legeza}.
The bipartite entanglement in systems of atomic Bose-Einstein
condensate was studied \cite{Milburn,Tonel}. The entanglement
entropy in the antiferromagnetic Heisenberg $XX$ chain and Ising
model was investigated \cite{Vidal}. In isotropic antiferromagnetic
Heisenberg model, the universal form of the entropy is predicted by
\cite{Latorre}
\begin{equation}
\label{eq1}S_L=\frac{c}{3}log_2L+k,
\end{equation}
where $c$ is the central charge and $k$ is a non-universal constant.
For a spin chain of open boundary condition, the analogous formula
$c/3$ should be replaced by $c/6$ for a part of length $L$ in an
infinite one-dimension system \cite{Holzhey,Calabrese}. The
analogous formula is dependent on the boundary conditions of the
block and the rest of the chain. If the block has two boundaries
with the rest of the chain then the factor is $c/3$, while if the
block has just one boundary as in the case of a block consisting of
the first adjacent spins of a semi-infinite chain, then the factor
is $c/6$. Recently, it is shown that a feeble central bound defect
\cite{Zhao} or single impurity in the boundary \cite{Es1} has strong
influence on the entropy, though the entropy measures the mutual
coupling of the two parts of a system in wave function. A weak
transverse boundary magnetic field impurity \cite{Zhou} and domain
walls \cite{Ren} generated by antiparallel magnetic field have
different effects on the entanglement entropy. Moreover, impurities
show their strong influences on spin correlation function
\cite{Lou,Fath}. It would be interesting to investigate the effects
of boundary impurities on the entanglement entropy in an opened
antiferromagnetic Heisenberg spin chain.

In this paper, the entanglement entropy of a spin$-1/2$
antiferromagnetic Heisenberg chain with boundary impurities located
at two ends is investigated. In scetion II, the Hamiltonian of an
antiferromagnetic Heisenberg spin$-1/2$ chain is presented. By using
the method of density-matrix renormalization-group (DMRG)
\cite{white,U}, the entropy of ground state is calculated and the
effects of the impurities are analyzed in section III. A discussion
concludes the paper.

\section{Hamiltonian of Heisenberg Spin Chain}

The Hamiltonian of a spin$-1/2$ Heisenberg opened chain with
boundary impurities at two ends can be written as
\begin{equation}
\label{eq2} H
=\sum_{i=2}^{N-2}J\vec{S}_i\vec{S}_{i+1}+\alpha(\vec{S}_1\vec{S}_2+\vec{S}_{N-1}\vec{S}_{N}).
\end{equation}
Where the coupling exchange $J>0$ corresponds to the
antiferromagnetic case, $\vec{S}_j$ are spin operators, $N$ is the
length of the spin chain. The coupling exchange $\alpha$ is impurity
interaction. For simplicity, $J=1$ is assumed in this paper.

The entropy is used as a measure of the bipartite entanglement. If
$|Gs\rangle$ is the ground state of a chain of N qubits, a reduced
density matrix of $L$ contiguous qubits can be written as
\begin{equation}
\label{eq3}\rho_L=Tr_{N-L}|Gs\rangle\langle Gs|.
\end{equation}
The bipartite entanglement between the right-hand $L$ contiguous
qubits and the rest of the system can be measured by the entropy
\begin{equation}
\label{eq4}S_L=-Tr(\rho_L\log_2\rho_L).
\end{equation}
One of the properties of the entropy of a block of the system can be
given by
\begin{equation}
\label{eq5}S_L=S_{N-L},
\end{equation}
since the spectrum of the reduced density matrix $\rho_L$ is the
same as that of $\rho_{N-L}$.

\section{Entanglement of Entropy with Impurities}

In order to calculate the entropy accurately using the method of
DMRG, the length of the spin chain needs to be relatively long. The
length of the spin chain is chosen to be $N=256$. The total number
of the density matrix eigenstates held in the system block is
$m=128$ in the basis truncation procedure.

To check the accuracy of the results from the method of DMRG, the
open boundary condition without impurities of $\alpha=1$ is
considered. The corresponding results of finite spin chain, which is
predicted by conformal field theory(CFT), can be considered as a
benchmark. It can be written as
\begin{equation}
\label{eq6}S_L=\frac{c}{6}\log_2[\frac{N}{\pi}\sin(\frac{\pi}{N}L)]+A,
\end{equation}
where $c$ is the central charge and $A$ is a non-universal constant
\cite{Chiara,Nicolas}. There are large oscillations between even and
odd $L$-value entropy. To avoid these relatively large oscillations,
the even-value entropy is chosen. The entropy $S_L$ between
contiguous $L$ qubits and the remain $N-L$ qubits is plotted as a
function of the subsystem L in Fig. 1 when $N=160, 200$ and $256$.
For $L<8$, the results of DMRG are slightly lower than that of CFT.
For $L>8$, almost perfect agreement between the two results is
obtained. For different values of $N$ with large $L$, $S_L$ is small
for small values of $N$. For small $L$, there is almost no
difference between different values of $N$. It seems that $S_L$
approaching a constant for very large $L$ is mainly due to finite
size effect. The entropy $S_L$ is also plotted as a function of
$\log_2[\frac{N}{\pi}\sin(\frac{\pi}{N}L)]$ in the inset of Fig. 1.
It is shown that the entropy appears as a straight line whose slope
is very close to $c/6$.

The entanglement entropy $S_{L}$ is plotted as a function of the
subsystem length $L$ for different values of the impurity
interaction $\alpha$ in Fig. 2(a). It is seen that the entropy $S_L$
increases with the subsystem length $L$ and then approaches a
constant when $L$ is very large for $\alpha=0.1, 2.0$. When
$\alpha=0.3, 0.5$, the entropy $S_L$ decreases slightly, then
increases and approaches a constant for very large $L$. The minimal
value is $1.53$ at $L=6$ for $\alpha=0.3$ and $1.16$ at $L=4$ for
$\alpha=0.5$. When $\alpha=0.1$, $S_L$ approaches the value about
$2.4$ for very large $L$. While for $\alpha=0.3, 0.5, 2.0$, $S_L$
approaches the value about $1.6$ for very large $L$. The influence
of the impurity at two ends of the Heisenberg spin$-1/2$ chain
depends on the value of the impurity interaction $\alpha$. For
$\alpha=\alpha_0=0.235$, the strength alternation of even bond and
odd bond in the center of the spin chain is minimized to almost
close to zero \cite{white}. For $\alpha<\alpha_0$, the even
sublattice is favored. This induces larger value of $S_L$. While for
$\alpha>\alpha_0$, the odd sublattice is favored. This induces
smaller value of $S_L$. If the value of $\alpha$ is less than
$\alpha_0=0.235$, the effects of the impurity on the entanglement
entropy $S_L$ is stronger. Therefore, the entanglement entropy $S_L$
of $\alpha=0.1$ is much larger than the $S_L$ of $\alpha=0.3, 0.5$
and $2.0$ \cite{Zhao,Es1,white}. The entropy $S_L$ is also plotted
as a function of $\log_2[\frac{N}{\pi}\sin(\frac{\pi}{N}L)]$ in the
inset of Fig. 2(a). It is seen that $S_L$ is almost a straight as a
function of $\log_2[\frac{N}{\pi}\sin(\frac{\pi}{N}L)]$ for very
large $L$.

If $S_{L\alpha}$ is the entropy with impurities at two ends and
$S_{L0}$ is the entropy without impurities, the difference of the
entropy $\Delta S_L$ can be defined as
\begin{equation}
\label{eq7}\Delta S_L=S_{L\alpha}-S_{L0}.
\end{equation}
The entropy difference $\Delta S_{L}$ may also be called "impurity
entanglement entropy" that is induced by adding impurities at two
ends of the spin chain \cite{Es1}. The entropy difference $\Delta
S_{L}$ is plotted as a function of the subsystem length $L$ for
different values of the impurity interaction $\alpha$ in Fig. 2(b).
It is shown that $\Delta S_L$ decreases and then approaches a
constant when the subsystem $L$ increases for $\alpha=0.1, 0.3,
0.5$. The value of $\Delta S_L$ increases and then approaches a
constant with increase of the subsystem length $L$ when
$\alpha=2.0$. It seems that the effect of impurity decreases with
the increase of the subsystem $L$ when $\alpha<\alpha_0=1.0$.
However, the effect of the impurity increases when
$\alpha>\alpha_0=1.0$. The entropy difference $\Delta S_{L}$ is also
plotted as a function of $\log_2[\frac{N}{\pi}\sin(\frac{\pi}{N}L)]$
in the inset of Fig. 2(b). It is seen that $\Delta S_L$ is almost a
straight as a function of
$\log_2[\frac{N}{\pi}\sin(\frac{\pi}{N}L)]$ for very large $L$.
Similar to that shown in Fig. 2(a), the entropy difference $\Delta
S_{L}$ of $\alpha=0.1$ is much larger than $\Delta S_L$ of
$\alpha=0.3, 0.5$ and $2.0$. This is mainly due to the fact that
small value of $\alpha<\alpha_0=0.235$ can induce stronger effect of
impurity on $\Delta S_{L}$ since the even sublattice is favored
\cite{Zhao,Es1,white}.

From Fig. 2, it is seen that both values of $S_L$ and $\Delta S_L$
are decreases when $\alpha$ increases especially for small number of
$L$. There are large differences of $S_L$ and $\Delta S_L$ for
different impurity interactions $\alpha$ when $L$ is small. If $L$
is quite large, $S_L$ and $\Delta S_L$ approach to constants. If
$\alpha=0.3, 0.5$ and $2.0$, $S_L$ approaches to about $1.6$ while
$\Delta S_L$ approaches to about zero for quite large $L$. It seems
that the effect of the impurity at two ends is very small for large
$L$. If $\alpha=0.1$, both $S_L$ and $\Delta S_L$ are quite large.
It seems that the small value of impurity interaction
$\alpha<\alpha_0=0.235$ can induce strong effect on $S_L$ and
$\Delta S_L$.

The central charge $c$ in Eq. (6) plays an important role in the
measurement of the entanglement entropy. The central charge $c$ can
be calculated numerically by \cite{Calabrese,Es1}
\begin{equation}
\label{eq8}c(L)=6[\frac{S_{L+2}-S_{L-2}}{T(L+2)-T(L-2)}],
\end{equation}
where
\begin{equation}
\label{eq9}T(L)=\log_2[\frac{N}{\pi}\sin(\frac{\pi}{N}L)].
\end{equation}
The central charge labeled by $c(L)$ is plotted in Fig. 3(a) as a
function of the subsystem length $L$ for different values of the
impurity interaction $\alpha$. When $\alpha=0.1$, the central charge
$c(L)$ increases to a peak and then decreases slowly with the
increase of the subsystem length $L$. The central charge $c(L)$
decreases and then approaches a constant with the increase of the
subsystem $L$ when $\alpha=2.0$. When $\alpha=0.3, 0.5$, the central
charge $c(L)$ increases and almost approaches a constant with the
increase of the subsystem length $L$. The central charges of
$c(L=6)$ and $c(L=4)$ are negative when $\alpha=0.3$ and $0.5$
respectively. This corresponds to the minimum values of $S_L$ shown
in Fig. 2(a). For $\alpha>0.235$, the central charge $c(L)$
increases with increasing value of $\alpha$. For $\alpha=0.1<0.235$,
c(L) of $\alpha=0.1$ is larger than that of $\alpha=0.3$, it
decreases and finally approaches to that of $\alpha=0.3$ when $L$ is
very large. The value of $c(L)$ of $\alpha=0.1$ is larger than that
of $\alpha=0.5$ if $L<20$. If $L>20$, $c(L)$ of $\alpha=0.1$ is
smaller than that of $\alpha=0.5$. This is mainly due to the
stronger singlet bonds on the even numbered links of the chain for
$\alpha<0.235$ \cite{white}. Since the central charge may clarify
the behavior of the entropy for large values of subsystem, the
central charge $c(L=80)$ is plotted as a function of impurity
interaction $\alpha$ in Fig. 2(b) when $1\ll L(=80)< N/2(=128)$. It
is seen that the central charge c reaches a minimum value when
impurity interaction $\alpha=0.235$. When impurity interaction
$\alpha<0.235$, the central charge $c$ decreases with impurity
interaction $\alpha$ increases. The central charge $c$ increases
with impurity interaction $\alpha$ increases when $\alpha>0.235$. It
approaches to $1.0$ when impurity interaction $\alpha$ is close to
$2.0$.

If the impurities are qutrit with spin-$1$ operators $\vec{S'}$, the
effects of qutrit-impurities on the entropy can also be
investigated. The entanglement entropy $S_L$ and the difference of
the entropy $\Delta S_{L}$ are plotted in Figs. 4(a) and 4(b)
respectively as a function of the subsystem length $L$ for impurity
interaction $\alpha$ and different impurities. The entropy $S_L$ and
the difference of the entropy $\Delta S_{L}$ are also plotted as a
function of $\log_2[\frac{N}{\pi}\sin(\frac{\pi}{N}L)]$ in the
insets of Figs. 4(a) and 4(b). Similar to that shown in Fig. 2, both
$S_L$ and $\Delta S_L$ of spin-$1$ decreases with the increase of
$\alpha$. The entropy $S_L$ increases and then almost approaches a
constant when the the subsystem length $L$ increases. The entropy of
impurity with spin-$1$ is much larger than that with spin-$1/2$. For
the impurity of spin-$1/2$, the entropies of $\alpha=0.5$ and
$\alpha=2.0$ are almost not distinguishable when the subsystem
length $L$ is very large. While for the impurity of spin-$1$, the
difference of the entropies of $\alpha=0.5$ and $\alpha=2.0$ are
quite large and approach a constant when $L$ is very large. It is
clear that the effects of qutrit impurity on the entanglement
entropy are much stronger than that of qubit impurity. It seems that
it is more easily to control the entropy of the system using qutrit
impurity.

\section{Discussion}

It is clear that the impurity interaction and the impurity spin have
a strong influence on the entanglement of the two subsystems
\cite{Es1,Wang1,E02}. For pairwise entanglement between the impurity
spin and the spin chain, the two boundary spins will have a strong
tendency to form a singlet pair when the impurity interaction is
large. This will reduce the entanglement between the boundary of the
two spin subsystems and the rest of the system. The value of
entanglement entropy is mainly determined by the density-matrix
spectra, extremely by the few largest eigenvalues of the reduced
density matrix \cite{U,Zhou,Zhao}. For qubit impurities, impurities
can affect the entropy between two subsystems by changing the
distribution of the reduced density-matrix spectra. If the
impurities are qutrits with the same impurity interaction, not only
the distribution of the reduced density-matrix spectra is changed,
but also the degree of the freedom of density-matrix spectra of the
subsystem is enlarged in the Hilbert space. This is similar to the
result of the entropy with the increase of subsystems
\cite{Latorre,Zhou,Ren}.

The effects of boundary impurities on the bipartite entanglement in
an antiferromagnetic Heisenberg open spin chain are discussed. Using
the method of density-matrix renormalization-group, entanglement
entropy is calculated for the even number subsystem. The
entanglement entropy decreases with the increase of the impurity
interaction while it increases with the increase of the subsystem
length. When the system length is very large, the entanglement
entropy approaches a constant due to finite size effect. The
influences of boundary impurities with qutrit of spin-$1$ are much
stronger than that of qubit of spin-$1/2$. With the same impurity
interaction, qutrit impurities can increase the entanglement. All
the results are dependent with the selection of even subsystem. This
is shown that the entropy of the system with qutrit impurity can be
more easily controlled.

\vskip 0.4 cm {\bf Acknowledgements}

It is a pleasure to thank Yinsheng Ling and Jianxing Fang for their
many helpful discussions. The financial supports from the National
Natural Science Foundation of China (Grant No. 10774108) and the
Creative Project for Doctors of Jiangsu Province of China are
gratefully acknowledged.

\clearpage
\newpage
\begin{figure}
\includegraphics[scale=0.6]{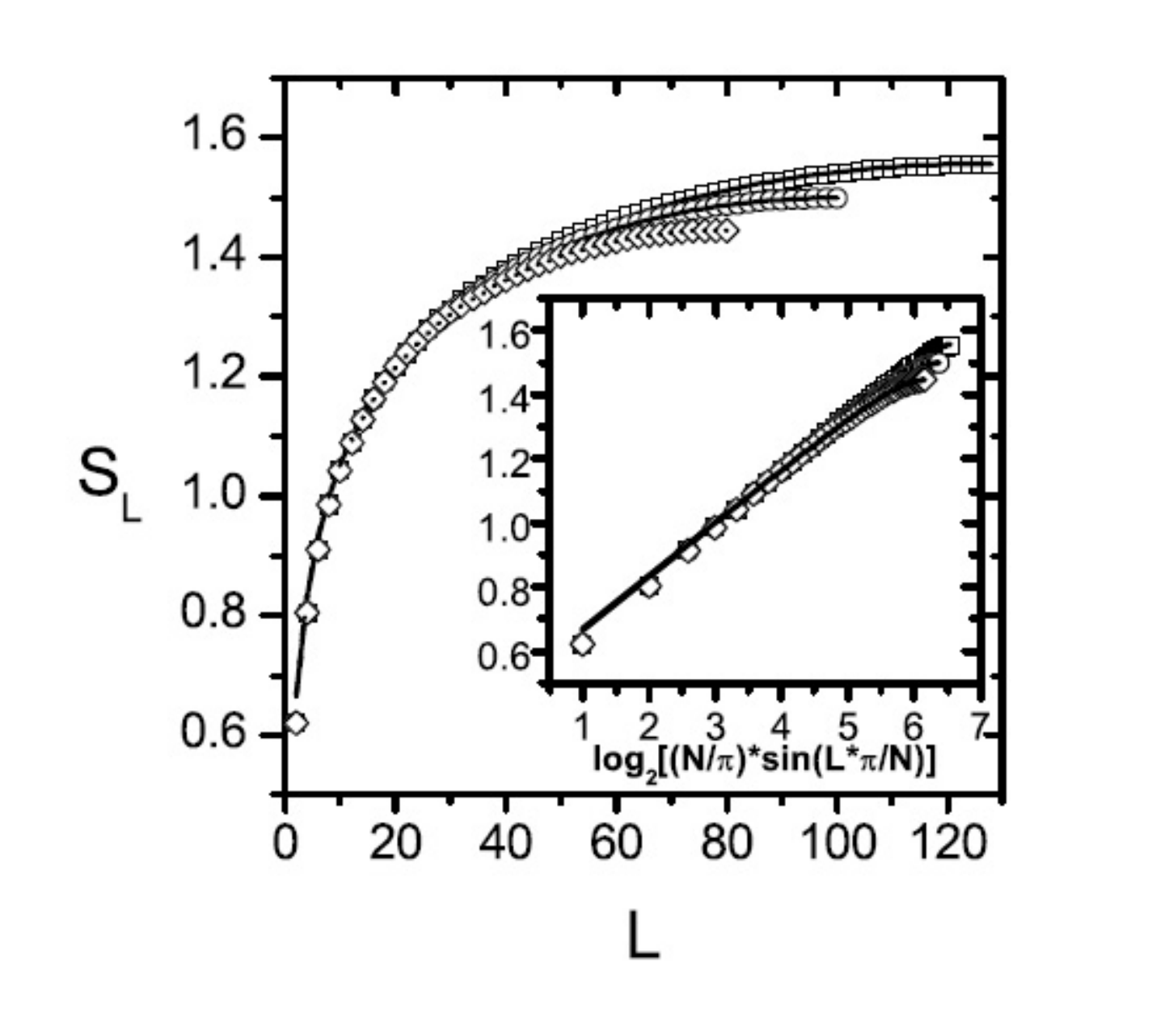}\caption{The entropy $S_L$ between contiguous $L$ qubits and the remain $N-L$
qubits is plotted as a function of the subsystem length $L$ for
$N=160, 200, 256$ (from bottom to top). The dashed line is obtained
from CFT and the symbol $\bigcirc$ is obtained from DMRG.}
\end{figure}

\clearpage
\newpage
\begin{figure}
\includegraphics[scale=0.6]{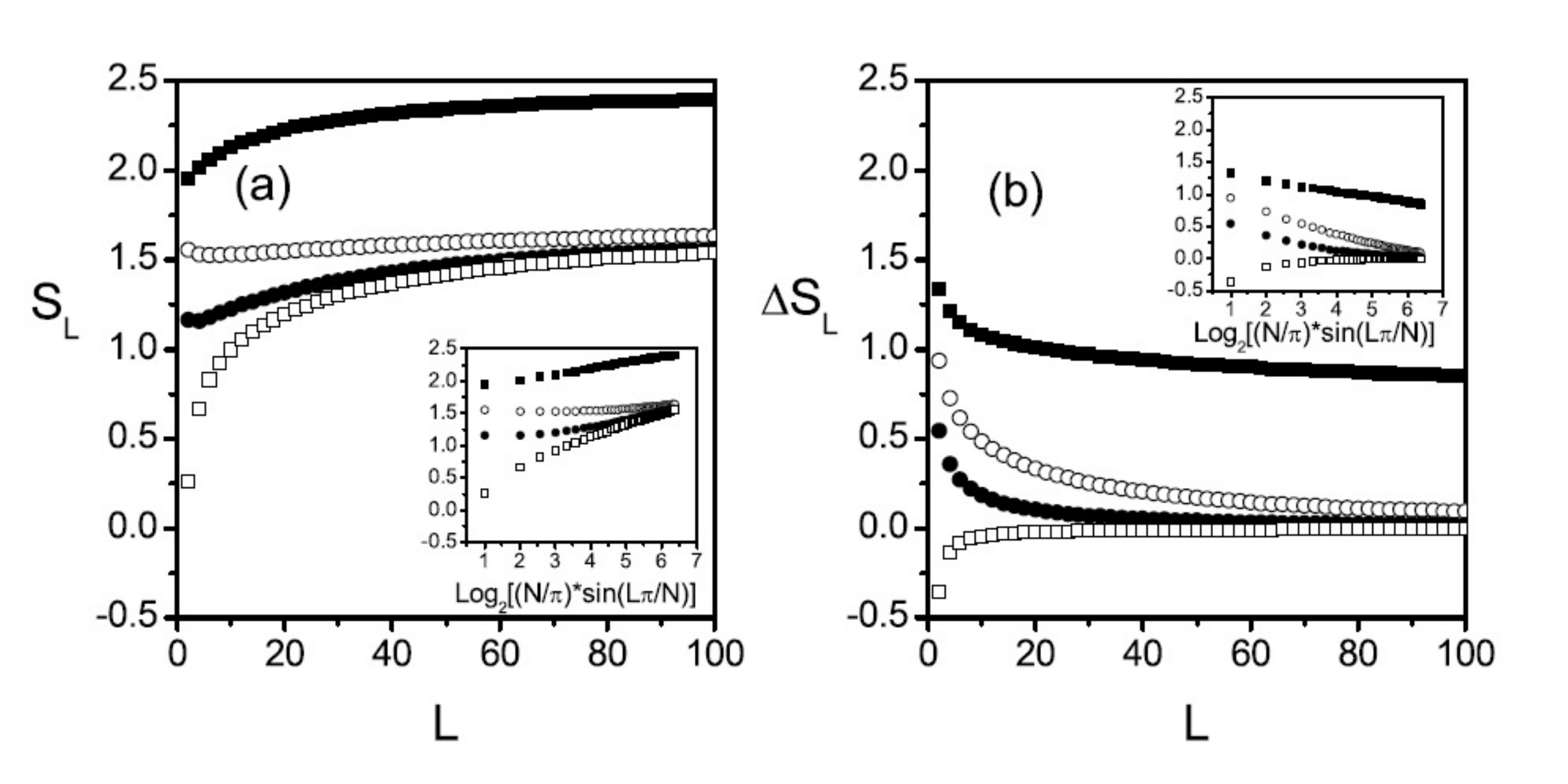}\caption{ The entanglement entropy $S_L$ and the entropy
difference $\Delta S_L$ are plotted as a function of the subsystem
length $L$ for different values of the impurity interaction
$\alpha$. (a). The entropy $S_L$. (b). The entropy difference
$\Delta S_L$. The symbols are for $\alpha=0.1(\blacksquare),
0.3(\circ), 0.5(\bullet), 2.0(\square)$.}
\end{figure}

\clearpage
\newpage
\begin{figure}
\includegraphics[scale=0.6]{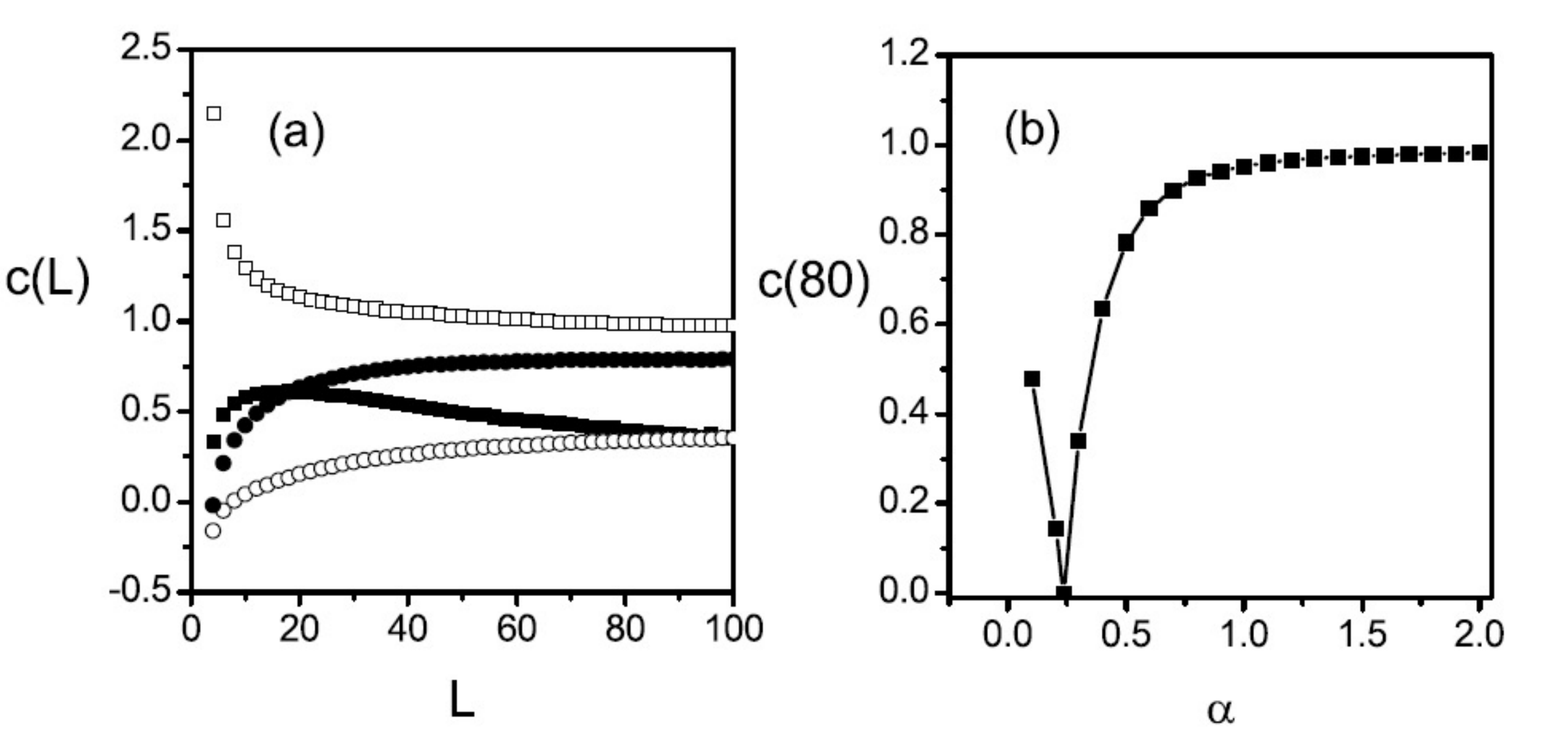}\caption{ (a). The central charge $c(L)$ is plotted as a function of the
subsystem length $L$ for different values of the impurity
interaction $\alpha$. The symbols are for $\alpha=0.1(\blacksquare),
0.3(\circ), 0.5(\bullet), 2.0(\square)$. (b). The central charge
$c(80)$ is plotted as a function of impurity interaction
$\alpha$.}
\end{figure}

\clearpage
\newpage
\begin{figure}[t]
\includegraphics[scale=0.6]{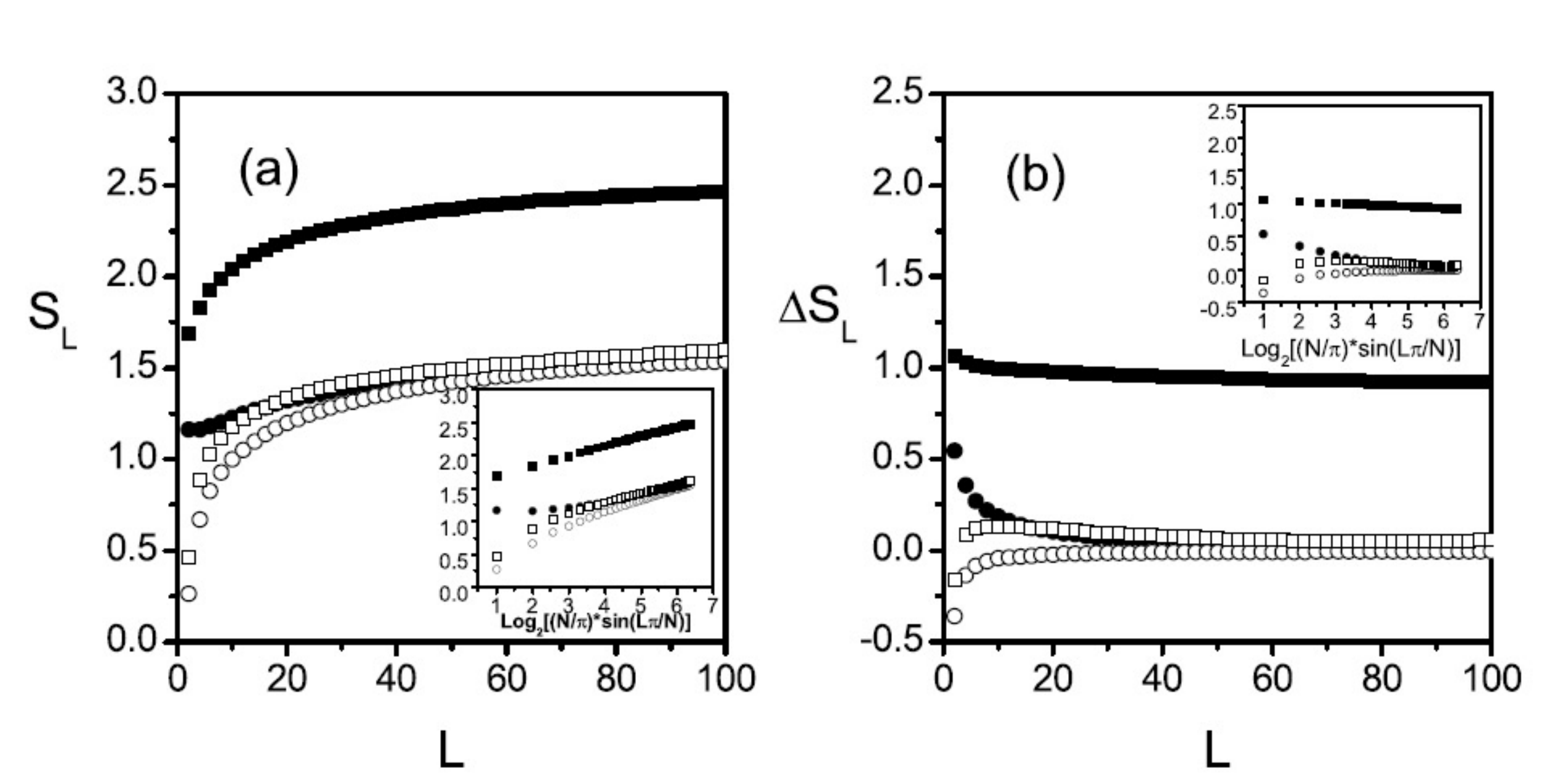}\caption{The entanglement entropy $S_L$ and the entropy
difference $\Delta S_L$ are plotted as a function of the subsystem
length $L$ for different values of the impurity interaction $\alpha$
and the spins $1/2$ and $1$. (a). The entropy $S_L$. (b). The
entropy difference $\Delta S_L$. The symbols are for qubit impurity
with $\alpha=0.5(\bullet), 2.0(\circ)$, qutrit impurity with
$\alpha=0.5(\blacksquare), 2.0(\square)$.  }
\end{figure}

\end{document}